\documentclass[journal,onecolumn]{IEEEtran}
\usepackage{mathtools}
\usepackage{cite}
\usepackage{color}
\usepackage{graphicx}
\usepackage{dcolumn}
\usepackage{bm}
\usepackage{amsmath,amsfonts}
\usepackage{float}
\usepackage[utf8]{inputenc}

	\newcommand \vecPhi{\boldsymbol \Phi}

\begin{document}
	\title{Optimal Actuation of Flagellar Magnetic  Micro-Swimmers}
\author{Yacine El Alaoui-Faris$^{1,2}$,
        Jean-Baptiste Pomet$^{1}$,
	Stéphane Régnier$^{2}$,
        and Laetitia Giraldi$^{1}$

      \thanks{$^{1}$ Université Côte d’Azur, Inria, CNRS, LJAD, McTAO team, Nice, France}
       \thanks{$^{2}$ Sorbonne Universités, CNRS, ISIR, Paris, France}
}

 \maketitle

\date{\today}

\begin{abstract}
We present an automated procedure for the design of optimal actuation for flagellar magnetic microswimmers based on numerical optimization. Using this method, a new magnetic actuation method is provided which allows these devices to swim significantly faster compared to the usual sinusoidal actuation. This leads to a novel swimming strategy which makes the swimmer perform a 3D figure-8 trajectory. This shows that a faster propulsion is obtained when the swimmer is allowed to go out-of-plane. This approach is experimentally validated on a scaled-up flexible swimmer.
\end{abstract}
\maketitle
\section{Introduction}
Untethered robotic micro-swimmers have the potential to make a serious impact on the development of new promising therapeutic techniques  \cite{NelsonKaliakatsos10,sitti2015biomedical,qiu2015magnetic,patra2013intelligent,mack2001minimally,fusco2014integrated,Lauga1}. However, due to their microscopic size, controlling their navigation through  fluids inside the body faces numerous challenges \cite{Lauga_2009,QiuLee14}. These devices can take different shapes and designs (helicoidal \cite{TabakYesilyurt14,Helicoidal_robot,QiuFujita15}, beating flexible tail \cite{dreyfus2005microscopic,KhalilDijkslag14},  2-linked structure \cite{JangAho18},  ciliary micro-robots \cite{kim2016fabrication}, etc.), and are actuated using multiple methods like chemical fuel propulsion \cite{MaJang16,JangHong17}, acoustic-based actuation \cite{AhmedBaasch16}, or external magnetic fields \cite{Xu2015,NelsonKaliakatsos10}. 
In this paper, we focus on sperm cell-inspired robotic swimmers that are composed of a magnetic head and an elastic tail and actuated by external magnetic fields \cite{KhalilDijkslag14,oulmas20173d}. As far as we know, the most commonly used actuation method for these swimmers is sinusoidal actuation. This consists in applying the superposition of a static orientating magnetic field parallel to the desired swimming direction and a perpendicular sinusoidal field that induces a planar symmetric beating of the tail, allowing a displacement along the swimming direction (\cite{abbott2009should}).
The purpose of this paper is to propose an automatic design method for the actuation of flexible magnetic micro-robots that optimizes the swimming speed. The motivation of this work is twofold : Firstly, we aim to provide a more efficient alternative  to the commonly used sinusoidal actuation, secondly, we provide a control design method than could be useful when dealing with more demanding tasks (where sinusoidal actuation fails), like swimming in a complex environment. Our method is based on the resolution of  an optimal control problem under the constraints of an approximate dynamical model of the swimmer's displacement. In particular, we focus on computing a magnetic field shape that  maximizes the horizontal swimming speed of these devices. This actuation pattern is experimentally validated on a scaled-up flexible magnetic swimmer and compared to the classical sinusoidal actuation.  
The key for this approach is the use of a dynamical model that is sufficiently computationally inexpensive to be used as a constraint for an optimization process but still accurate enough to fit the experiments. For this purpose, we use a 3D simplified dynamic model  generalizing the planar models of  \cite{moreau2018asymptotic,alouges2013self,alouges2015can} that takes into account the magnetic effect, the elasticity of the tail and the hydrodynamic effects which are simplified by the Resistive Force Theory \cite{gray1955propulsion,friedrich2010high}. Using this dynamic model as a constraint, the optimal control problem is numerically solved with a direct method using the software  \textit{ICLOCS} \cite{falugi2011imperial}.
This results in a magnetic actuation pattern that allows a significantly faster propulsion than the common sinusoidal actuation. The optimal magnetic field has two time-varying components which makes the trajectory of the swimmer non planar. In particular, the swimming strategy under this new magnetic field makes the swimmer perform a 3D 'figure-8' trajectory around the swimming direction. This indicates the importance of going outof plane in order to maximize the propulsion speed of flexible swimmers. 
The optimal actuation is experimentally validated on a scaled-up flexible magnetic swimmer using a similar setting as \cite{oulmas20173d} and compared to the sinusoidal field. The horizontal displacement of the experimental swimmer is accurately predicted by the dynamic model for both fields. These results show the usefulness of the simplified swimmer model for the design of optimal controls that are usable in experiments.    
~\\

\section{Experimental Setting} 
\begin{figure}[h!]
\begin{center}
\includegraphics[scale=0.5]{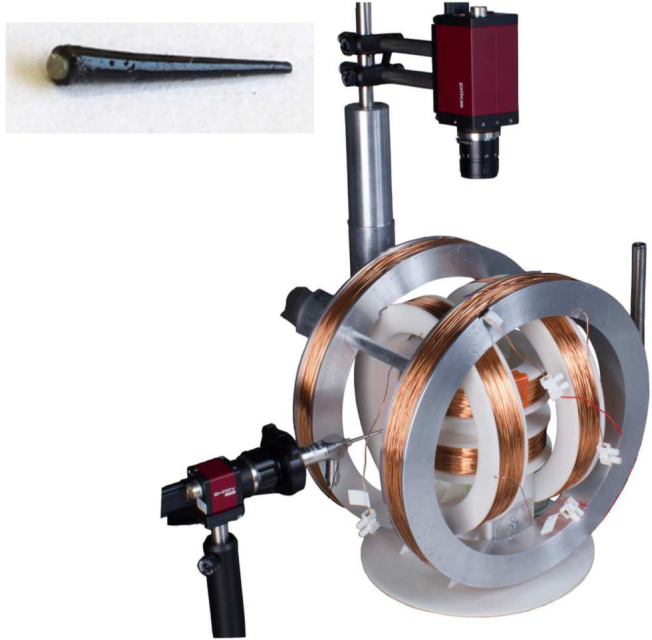}
\end{center}
\caption{Experimental setup (\cite{oulmas20173d}). The three orthogonal Helmholtz coils generate a homogeneous magnetic field in the center, where the swimmer has been placed. The swimmer is tracked using two perpendicular cameras.    Shown in the corner :  Scaled-up flexible magnetic swimmer used in the experiments. The tail is made of an elastomer shaped by a 3D-printed mold. The head is a  magnetic disk. \label{Fig:Robot}}
\end{figure}

~ \\
The flexible swimmer used consists of a $0.3 mm$ height, $0.77 mm$ diameter magnetic disk (Neodyminum-Iron-Boron permanent magnet)  attached to a silicone tail with a length of $7 mm$ and a diameter of $1 mm$ (see Figure \ref{Fig:Robot}). The swimmer is immersed in pure glycerol to ensure low-Reynolds conditions ($\approx 10^{-2})$. It is placed at the center of three orthogonal Helmholtz coil pairs that generate the actuating magnetic field.  Each pair of coils is driven by a servoamplifier (Maxon Motor) which outputs a constant current for a fixed input voltage.  Two cameras provide a side view and a top view of the swimmer, and its position in 3D is tracked in real-time using Visp software \cite{Marchand05b}. More information on the experimental setting used can be found in \cite{oulmas20173d,xu2014propulsion}. \\
We consistently use a fixed frame $(x,y,z)$ where the $z$ axis is vertical and the $y$ axis is along the axis of the side camera. In what follows , the desired swimming direction will be along the $x$-axis. 

The propulsion of the experimental swimmer is characterized by measuring its velocity-frequency response (given by the dotted line in Figure (\ref{Fig:Fitting}))  under a magnetic field of the form :
\begin{equation}
\boldsymbol{B}(t) = \begin{pmatrix} B_x & B_y\sin(2\pi f t) & 0 \end{pmatrix}^T ,
\end{equation}
where $B_x = 2.5 mT$,$B_y = 10mT$. \\

\section{Modeling of the swimmer}
A simplified dynamic model of the swimmer is used to simulate the swimmer's displacement under a  three-dimensional magnetic field. The hydrodynamic effects on the swimmer are approximated by Resistive Force Theory \cite{gray1955propulsion} and the shape of the tail is discretized into an articulated chain of $N$ rigid slender rods, generalizing the planar swimmer models of \cite{moreau2018asymptotic,alouges2013self,alouges2015can}.

\subsection{Kinematics of the swimmer}
\begin{figure}[H]
\includegraphics[width=0.9\textwidth]{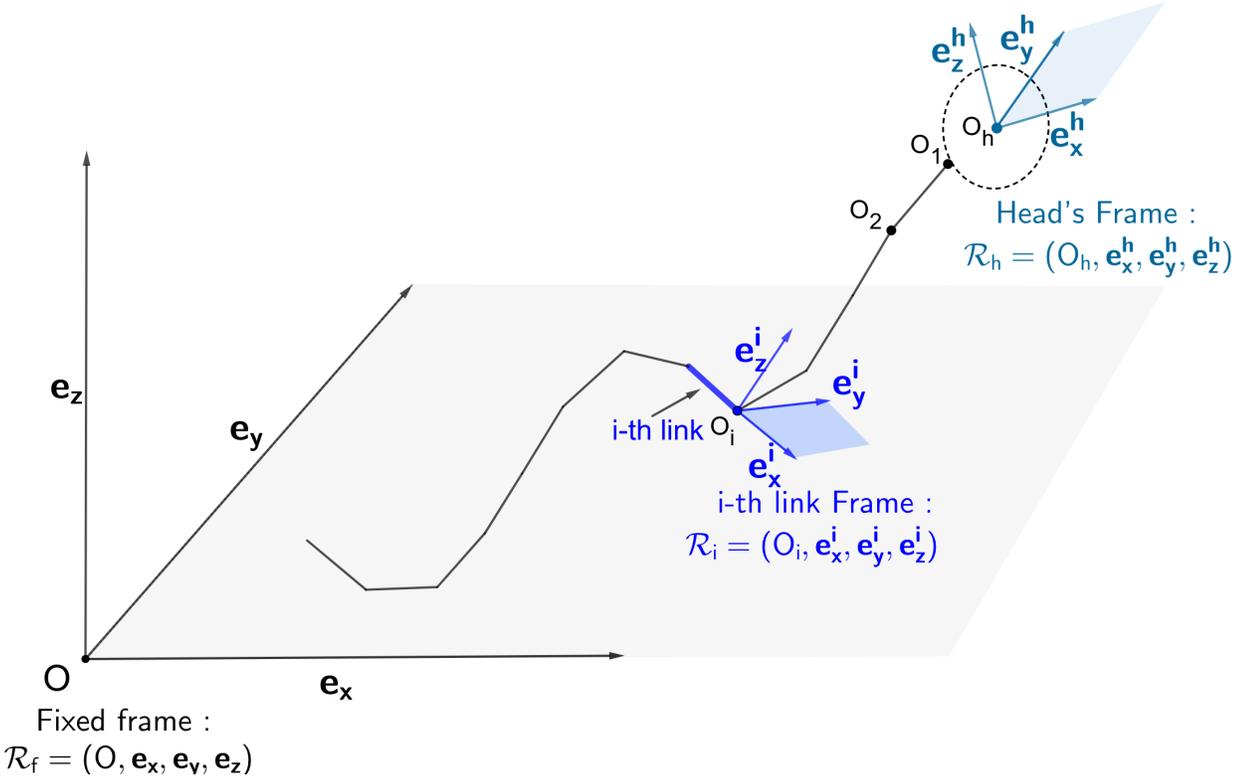}
\caption{Reference and local frames of the discrete-shape model. The swimmer's head frame is oriented relative to the reference frame. For each link $i$, the corresponding local frame $\mathcal{R}_i$ is oriented relative to $\mathcal{R}_{h}$.} 
\label{frames}
\end{figure}

We assimiliate the orientation and position of the swimmer with the orientation and position of the head, to  which we associate the moving frame $\mathcal{R}_{head} = (O_h,\boldsymbol{e_x^h},\boldsymbol{e_y^h},\boldsymbol{e_z^h})$, where $O_h$ is the center of the head. The orientation of each link $i$ is represented by  the moving frame $\mathcal{R}_i  = (O_i,\boldsymbol{e_x^i},\boldsymbol{e_y^i},\boldsymbol{e_z^i})$, where $O_i$ is the extremity of the $i$-th link. We call $R_{head} \in SO(3)$ the rotation matrix that allows the transformation of coordinates from the fixed reference frame to $\mathcal{R}_{head}$. Similarly, the matrix $R_i \in SO(3)$, for $i=1 \cdots N$, denotes the relative rotation matrix that transforms coordinates from $\mathcal{R}_{head}$ to $\mathcal{R}_i$. We use the angles $(\theta_x,\theta_y,\theta_z)$ resulting from a $(X-Y-Z)$ rotation sequence to parametrise $R_{head}$. Since each link is considered to be slender and axisymmetric, the vectors $\boldsymbol{e_y^i}$ and  $\boldsymbol{e_z^i}$ are only defined up to one rotation around, hence we parametrize each matrix $R_i$ by only two angles $(\phi^i_y,\phi^i_z)$, resulting from a $Y-Z$ rotation sequence relative to the head frame. \\ 
With these notations, the swimmer is described by two sets of variables : 
The $6$ \textit{Position variables}: $(\boldsymbol{X},\boldsymbol{\Theta})$ where 
\begin{equation}
 \begin{aligned}
  X  &= (x_h,y_h,z_h) \in \mathbb{R}^3 \text{and} \\ 
  \Theta &= (\theta_x,\theta_y,\theta_z) \in [0, \ 2\pi]^{3},
  \end{aligned}
\end{equation}, and the $2N$ \textit{Shape variables}, denoted by 
\begin{equation}
\vecPhi = (\phi^1_y,\phi^1_z,\cdots,\phi^N_y,\phi^N_z).
\end{equation}
\subsection{Dynamics of the swimmer}
We assume that the swimmer is immersed in an unbounded domain of a viscous fluid. Due to its scale, we consider that the swimmer moves at low Reynolds Number, i.e. that viscosity prevails over inertia, and thus that the second order terms in Navier-Stokes are small enough to be neglected, leading to Stokes equations. In what follows, the hydrodynamic interactions between the fluid and the swimmer are further simplified by using the Resistive Force Theory (RFT) framework \cite{gray1955propulsion}, where the interactions on the global scale between a slender Stokesian swimmer and the surrounding fluid are neglected in favor of the local anisotropic friction of the surface of the slender body with the nearby fluid. This results in explicit expressions of the density of force applied by the fluid to the swimmer that are linear with respect to the velocities, hence to $(\dot{\boldsymbol X}, \dot{\boldsymbol \Theta}, \dot{\boldsymbol \Phi})$. From the integration of the local hydrodynamic force densities on each link, we are able to obtain the expression of $\boldsymbol{F^h_{head}}$ and $\boldsymbol{F^h_i}$, the hydrodynamic drag forces on the head of the swimmer and on each link $i \in (1,\cdots,N)$ respectively. The expressions for $\boldsymbol{T^h_{head,P}}$ and $\boldsymbol{T^h_{i,P}}$, the moments of these forces about any given point $P$, are similarly derived from the local drag densities.
We consider that the head of the swimmer is magnetized along the $e^h_x$ axis. Denoting by  $\boldsymbol{M}$ the magnetization vector of the head and considering an external homogeneous time-varying field $\boldsymbol{B}(t)$, the following torque is applied to the swimmer:  
\begin{equation}
\boldsymbol{T^{mag}} = \boldsymbol{M} \times \boldsymbol{B}(t).
\end{equation}

The acceleration terms in the dynamics are neglected due to the Low Reynolds assumption \cite{yates1986microorganisms}, thus, the balance of forces and torques applied on the swimmer gives :  
\begin{equation}
\label{eqdyn_act}
\left\{
\begin{aligned}
 \boldsymbol{F^h_{head}} + \sum_{i=1}^N \boldsymbol{F^h_i} =& 0_3 \,, \\
 \boldsymbol{T^h_{head}} + \sum_{i=1}^N \boldsymbol{T^h_{i,H}} =& - \boldsymbol{T^{mag}} \,, \\
\end{aligned}
\right.
\end{equation} which leads to $6$ independent equations. In addition to these equations, we take the internal contributions of the tail  into account by adding  the balance of torque on each subsystem consisting of the chain formed by the links $i$ to $N$ for $i = 1\cdots N$. These additional  $3N$ equations reduce to $2N$ non-trivial equations by taking only the components perpendicular to the link $k$ when calculating the sum of the torques from $k$ to $N$. The elasticity of the tail is discretized by considering a restoring elastic moment $\boldsymbol{T^{el}_i}$ at each joint $O_i$ that tends to align each pair $(i,i+1)$ of adjacent links with each other: 
\begin{equation}
\boldsymbol{T^{el}_i} = k_{el} \boldsymbol{e_x^i} \times \boldsymbol{e_x^{i-1}}.
\end{equation}
Thus, the dynamics of a flexible micro-swimmer with a passive tail are described the following  system of  $2N + 6$ equations : 
\begin{equation}
\left\{
\begin{aligned}
\boldsymbol{F^h_{head}} + \sum_{i=1}^N \boldsymbol{F^h_i}  &= O_3 \,,  \\
 \boldsymbol{T^h_{head}} + \sum_{i=1}^N \boldsymbol{T^h_{i,H}} &= - \boldsymbol{T^{mag}} \,,  \\
  \sum_{i=1}^N \boldsymbol{T^h_{i,1}} . \boldsymbol{e_y^1}  &= - \boldsymbol{T^{el}_1}.\boldsymbol{e_y^1} \,,  \\ 
  \sum_{i=1}^N \boldsymbol{T^h_{i,1}} .  \boldsymbol{e_z^1} &= - \boldsymbol{T^{el}_1}.\boldsymbol{e_z^1} \,, \\ 
      \mathrel{\makebox[\widthof{=}]{\vdots}} \\ 
  \boldsymbol{T^h_{N,N}}   . \boldsymbol{e_y^n}        &= - \boldsymbol{T^{el}_n}.\boldsymbol{e_y^n} \,, \\ 
   \boldsymbol{T^h_{N,N}}   . \boldsymbol{e_z^n}       &= - \boldsymbol{T^{el}_n}.\boldsymbol{e_z^n} \,.
\end{aligned}
\right.               
\end{equation}

Following Resistive Force Theory, the hydrodynamic contributions (right-hand side of the previous system) are linear with respect to the rotational and translational velocities, thus, the previous system can be rewrited matricially in the form :  

\begin{equation}
 M^h(\boldsymbol \Theta,\boldsymbol \Phi)\begin{pmatrix} \boldsymbol{\dot{X}} \\ \boldsymbol{\dot{\Theta}} \\ \boldsymbol{\dot{\Phi}}\end{pmatrix} = {B(\boldsymbol X, \boldsymbol \Theta,\boldsymbol \Phi)} \,,
\end{equation} where the left-hand side of the equation represents the hydrodynamic effects on the swimmer and the right-hand side $B(\boldsymbol X,\boldsymbol \Theta,\boldsymbol \Phi)$ corresponds the magnetic and elastic contributions on the swimmer. 
The previous equation can be rewritten as a control system, where the dynamics of the swimmer are affine with respect to the components of the actuating magnetic field viewed as a control: 
\begin{equation}
\label{Dynamique} 
\begin{pmatrix}\dot{\boldsymbol{X}} \\ \dot{\boldsymbol{\Theta}} \\\dot{\boldsymbol{\Phi}} \end{pmatrix} = \boldsymbol{F_0}(\boldsymbol{\Theta},\boldsymbol{\Phi}) +\begin{pmatrix} \boldsymbol{B_x}(t) & \boldsymbol{B_y}(t) & \boldsymbol{B_z}(t) \end{pmatrix} \begin{pmatrix} \boldsymbol{F_1}(\boldsymbol{\Theta},\boldsymbol{\Phi})\\ \boldsymbol{F_2}(\boldsymbol{\Theta},\boldsymbol{\Phi}) \\ \boldsymbol{F_3}(\boldsymbol{\Theta},\boldsymbol{\Phi}) \end{pmatrix} \,,
\end{equation}
where the vector fields $F_0,\cdots,F_3$ are functions of the columns of $(M^h)^{-1}$ and of the magnetic and elastic constants.
\subsection{Parameter Identification}  

\begin{figure}[H]
\includegraphics[width=0.9\textwidth]{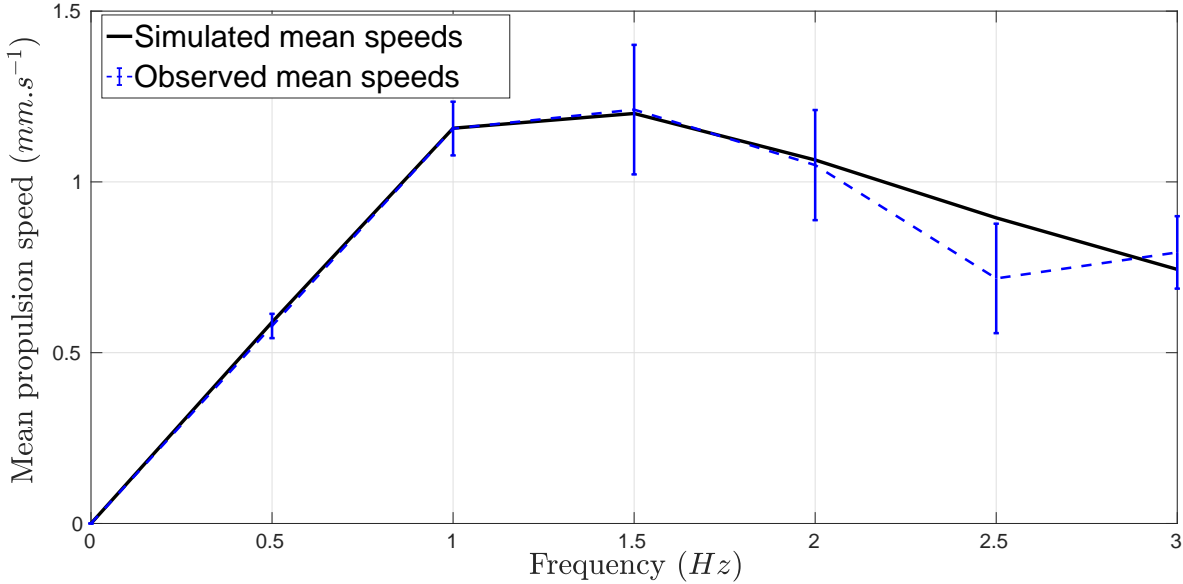}
\caption{Observed (N = 6)  and simulated horizontal swimming velocities for the frequency range $f = (0\cdots 3 Hz)$.  \label{Fig:Fitting}}
\end{figure}

The hydrodynamic parameters (RFT coefficients) and the elasticity coefficient of the model are identified by  non-linear fitting  to match the observed velocity-frequency response curve of the experimental swimmer under sinusoidal actuation. Three links were used for the approximation of the tail. Figure \ref{Fig:Fitting} shows the agreement between the experimental and simulated frequency responses. 

It is worth noting that using a finer discretization (more than $3$)  of the tail only marginally improves the fitting error while adding to the computational cost of the magnetic field optimization.

~\\ 
\section{Magnetic Field Optimization}
\begin{figure}[h]
\includegraphics[width=0.9\textwidth]{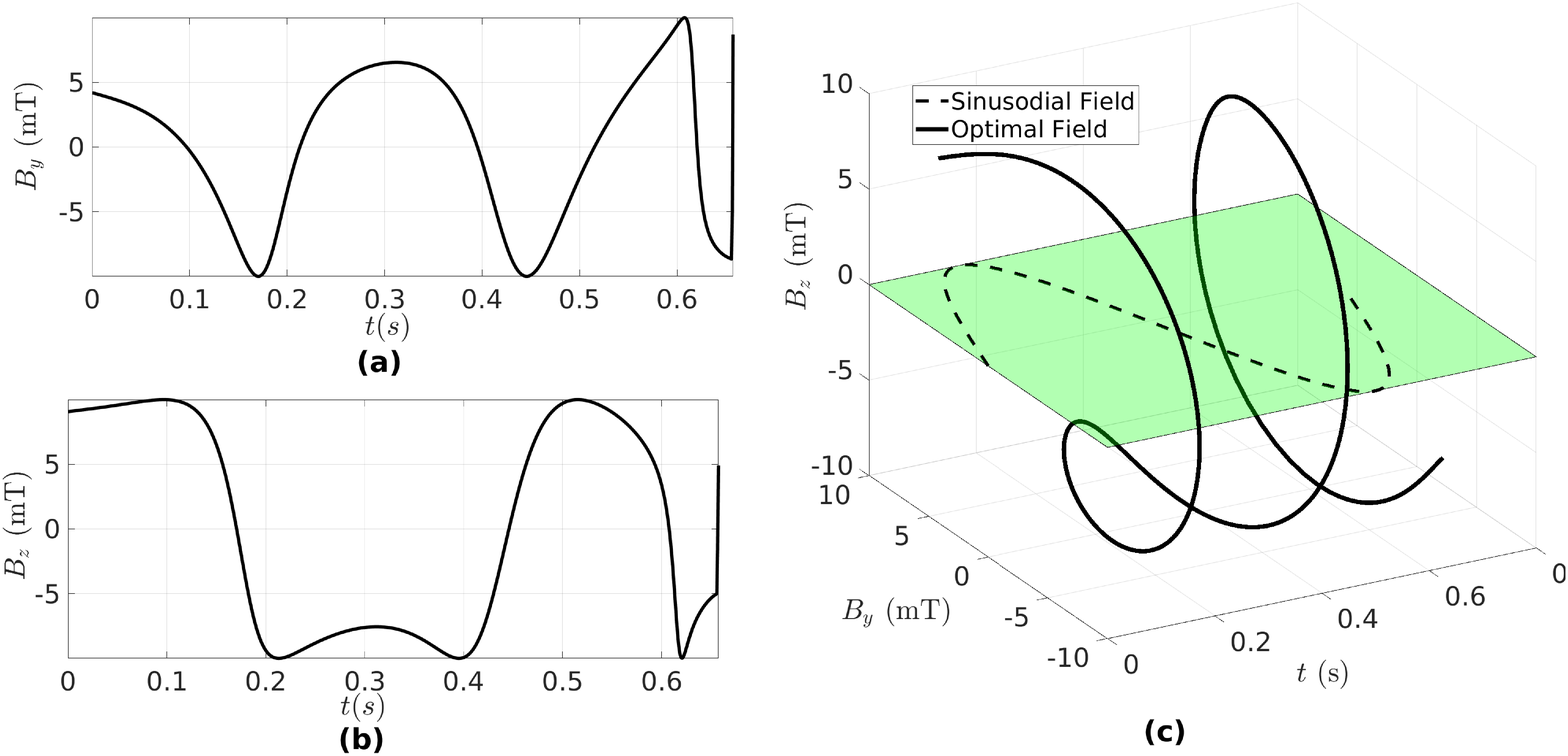}
\caption{Actuating magnetic field that maximizes the horizontal speed of the swimmer. $\boldsymbol{(a)}$ and $\boldsymbol{(b)}$ are respectively the $y$ and $z$ components of the magnetic field. $\boldsymbol{(c)}$ : Shape of the actuating \textit{optimal} magnetic field deriving from the optimization process during one period. \label{Fig_Optfields}}
\end{figure}

\begin{figure}[h]
\includegraphics[width=0.9\textwidth]{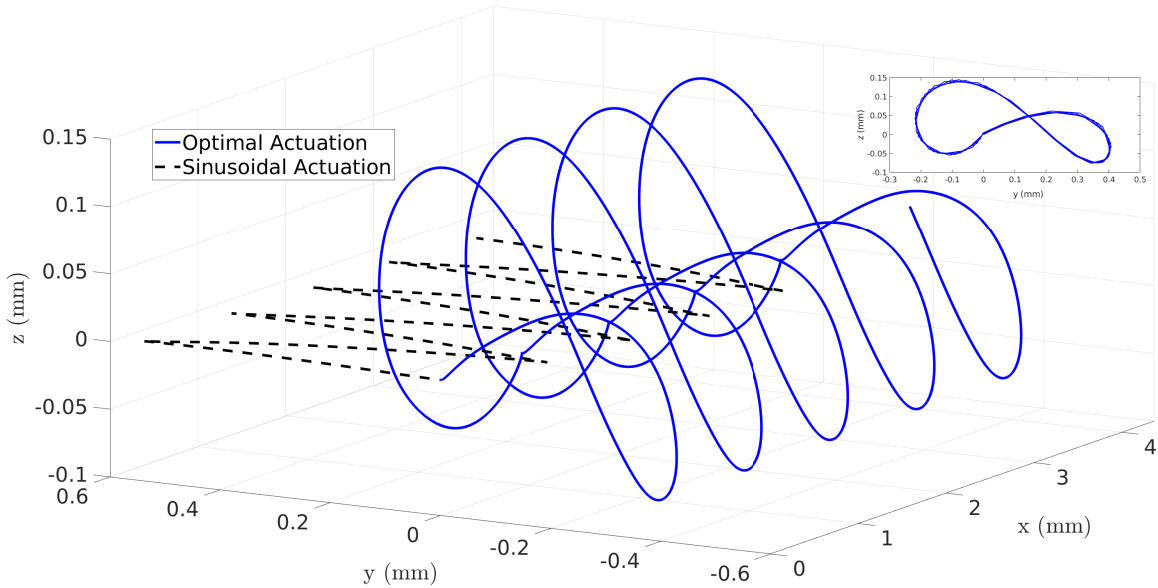}
\caption{Simulated trajectory of the head of the swimmer under both the optimal and sinusoidal field for 3 seconds of straight swimming under both actuation patterns. In the corner, the trajectory in the $y-z$ plane.\label{Displacement3D}}
\end{figure}

In what follows we focus on finding the time-varying magnetic field that maximizes the mean horizontal propulsion speed of the swimmer under the constraint of a periodic deformation. The admissible controls are of the form $(B_x,B_y(t),B_z(t))$,  where the static orientating field $B_x$ is fixed and the two dimensional time-varying actuating field ($B_y(t),B_z(t)$) is optimized. The same amplitude constraints as for the sinusoidal field are imposed ($B_x = 2.5 mT$ and the norm of $(B_y(t),B_z(t))$ is bounded to 10mT). 
The deformation period is chosen to be the same as the observed optimal period for a sinusoidal actuation and it is imposed that the swimmer returns in its initial orientation and position on the $y$ and $z$-axis at the end of the period.   
This optimal control problem is numerically solved with the direct solver ICLOCS \cite{falugi2011imperial}. 
Figure \ref{Fig_Optfields} shows the result of the numerical optimization, where we can see  the $y$ $\boldsymbol{(a)}$  and $z$ $\boldsymbol{(b)}$ components of the optimal magnetic field. Interestingly enough, this results in a simple  magnetic field shape, as shown in $\boldsymbol{(c)}$, where the actuating strategy over one deformation period is a partial rotation (about two thirds of a circle) of the magnetic field followed by a full rotation in the opposite direction. Repeating this pattern over time makes the swimmer revolve around the $x$ axis, drawing a "figure-$8$" in the $y-z$ plane. Figure \ref{Displacement3D} shows the simulated trajectory of the swimmer under this actuation strategy compared to the trajectory under the sinusoidal field.
The simulated fields perform better (mean propulsion speed of $1.45 mm.s^{-1}$) than the sinusoidal actuation (mean propulsion speed of $1.2 mmm.s^{-1} $). 

~\\
\section{Experimental Results}

\begin{figure}[h]
\includegraphics[width=0.9\textwidth]{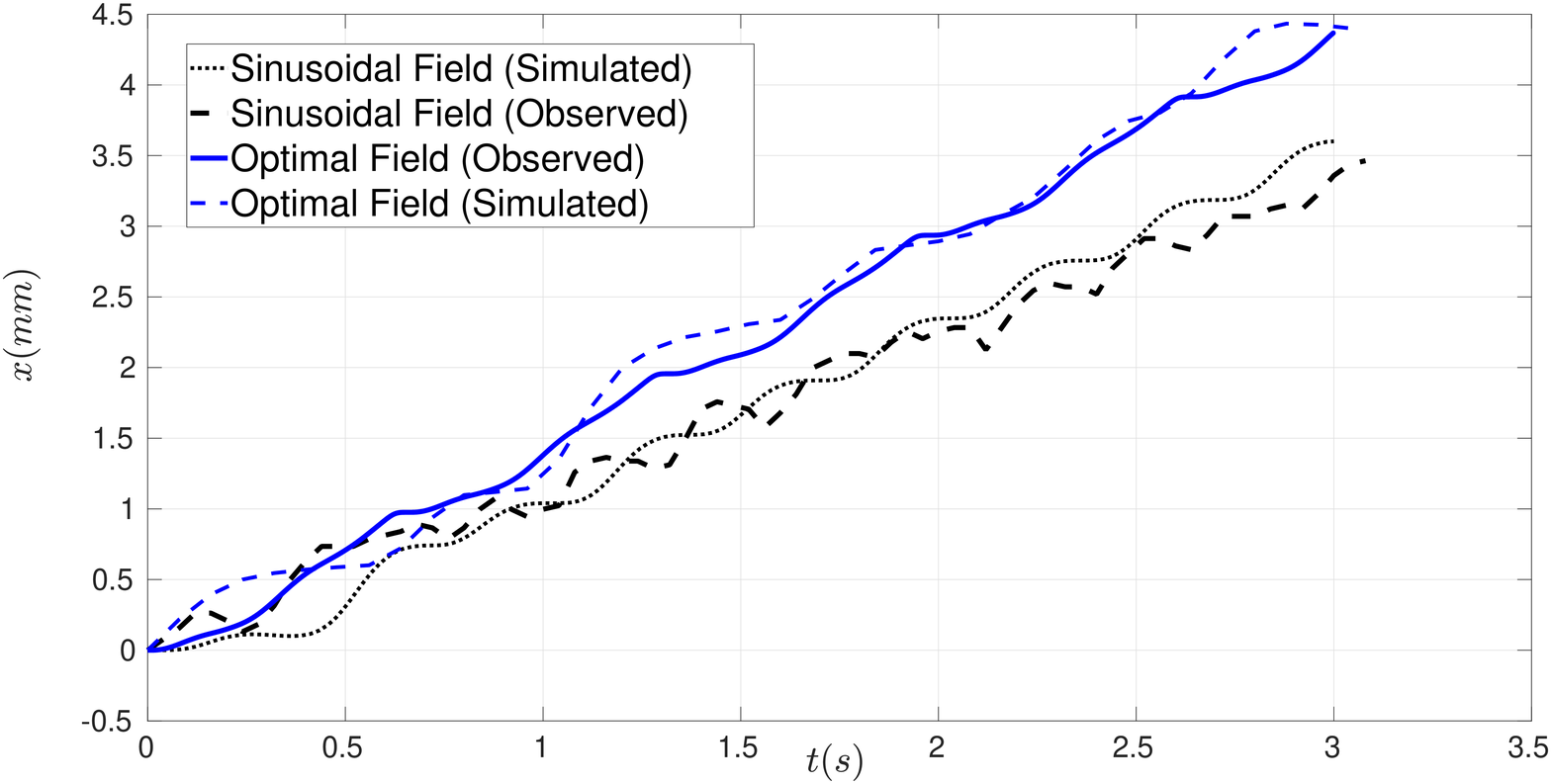}
\caption{Observed and Predicted horizontal displacements of the swimmer actuated by the optimal magnetic field and the sinusoidal magnetic field for 3 seconds of straight swimming under both actuation patterns. \label{Trajs_all}}
\end{figure}

\begin{figure}[h]
\includegraphics[width=0.9\textwidth]{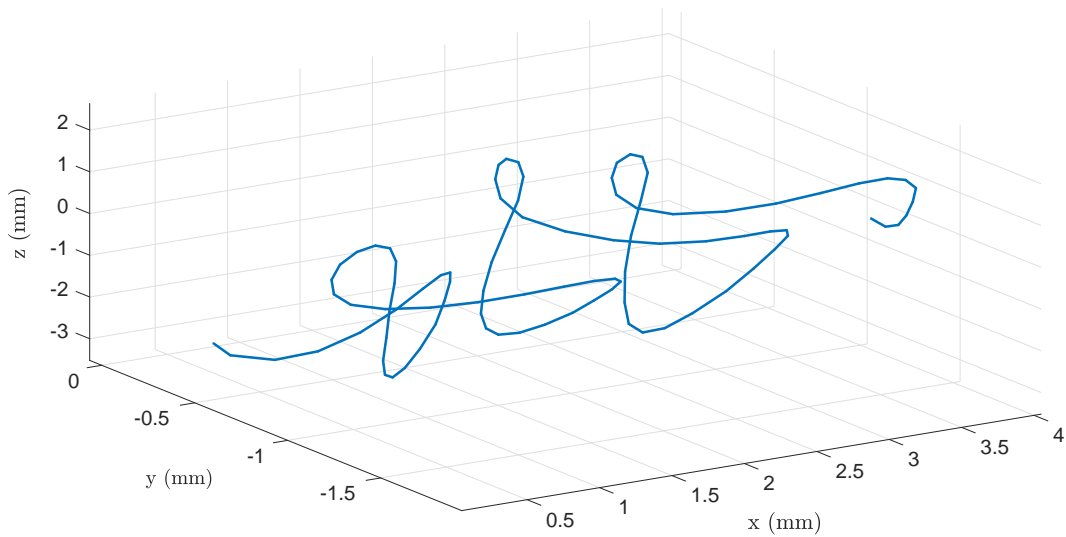}
\caption{Observed trajectory of the swimmer actuated by the optimal magnetic field given in Figure \ref{Fig_Optfields} during three periods of the magnetic field. \label{fig_trajectory}}
\end{figure}

\begin{figure}[h]
\includegraphics[scale=0.7]{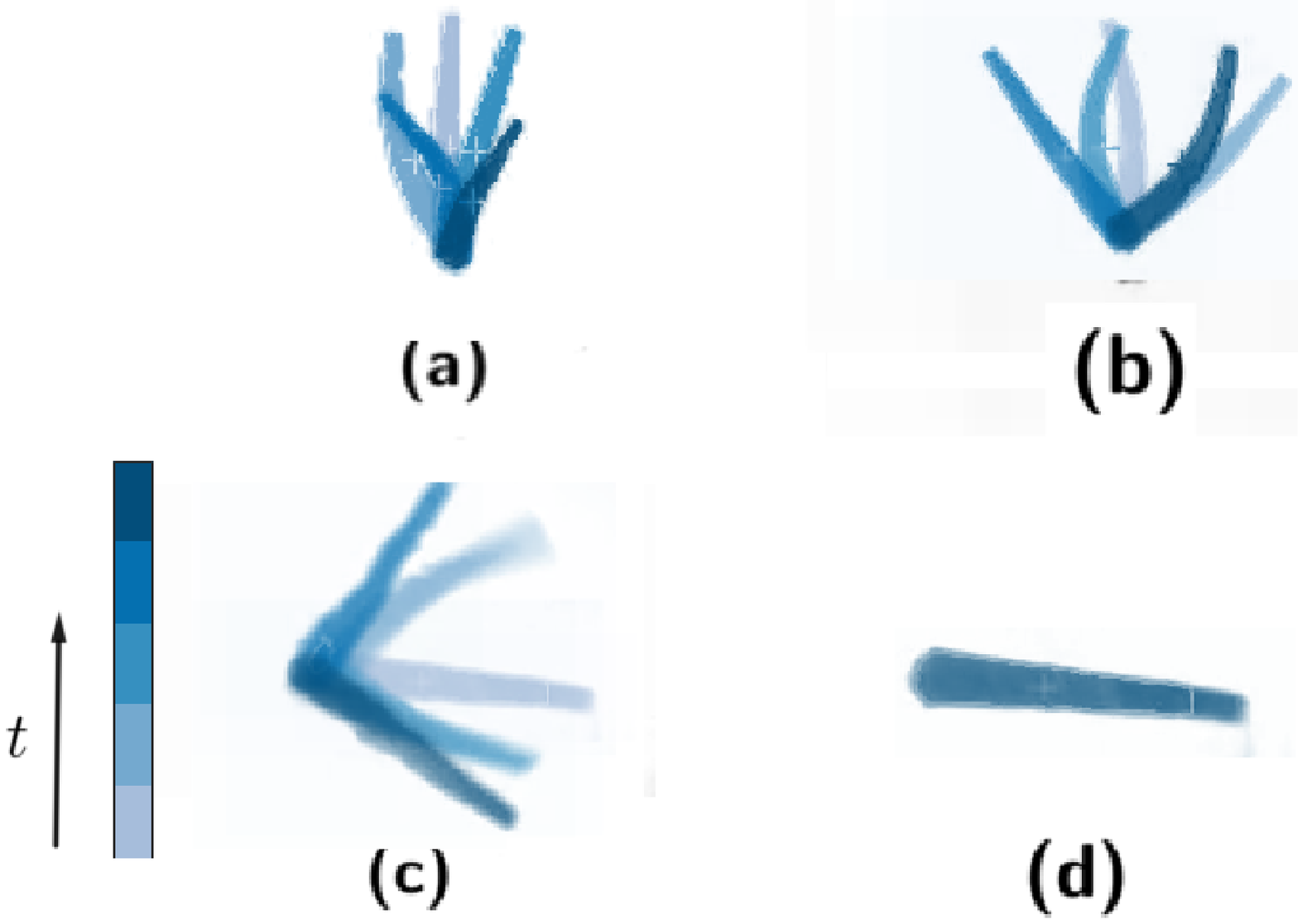}
\caption{Deformation pattern of the swimmer actuated by the optimal magnetic field and the sinusoidal magnetic field over one period in two planes. ( $\boldsymbol{(a)}$ : Optimal field,  top view, $\boldsymbol{(b)}$ : Sinusoidal field, top view, $\boldsymbol{(c)}$: Optimal field, side view, $\boldsymbol{(d)}$ : Sinusoidal field, side view.). The snapshots of the swimmer are taken at equal time steps over a period of the actuating fields. We observe (as expected) no deformation in the side plane for the sinusoidal field.
\label{fig:shape_exp}}
\end{figure}

The numerical solution of the optimal control problem (Figure \ref{Fig_Optfields}) is interpolated by its truncated Fourier expansion (first 10 modes) and then implemented in the magnetic generation system to actuate the swimmer. As predicted by the simulations, the optimal field out-performs the sinusoidal field in terms of horizontal propulsion speed, as shown in Figure \ref{Trajs_all}. Figure \ref{fig_trajectory} shows the 3D trajectory of the experimental swimmer during one period. Figure \ref{fig:shape_exp} shows the deformation pattern undergone by the swimmer in the two perpendicular planes. Under optimal actuation, the swimmer reaches a mean horizontal propulsion speed of $1.54 \pm 0.3  mm.s^{-1}$ ($N = 6$). The mean relative error ($\infty$-norm) between the simulated and observed $x$-displacement is $0.16 (\pm 0.02)$.
We refer the reader to the videos in the supplementary material for a side-by-side comparison of the displacements of the swimmer under both the optimal field and the sinusoidal field from two viewpoints (from the top and from the side).          
~
~

\section{Discussion}
~\\
The optimization process exploits two time varying components of the field in order to maximize the horizontal speed of the swimmer, which  allows the swimmer to perform a  3D trajectory. Interestingly enough, this new swimming strategy obtained by numerical optimization shows the necessity of allowing flagellar swimmers to go out-of plane in order to swim at a maximal propulsion speed, as illustrated in Figure \ref{Displacement3D} and Figure \ref{fig_trajectory}. The effectiveness of non-planar actuation has been  corroborated in the literature by studies where non-planar helical waves have  been shown to induce a faster propulsion speed for flagellar swimmers. For example, in \cite{khalil2018controllable}, the sperm-like swimmer's swimming speed increases between 1.2 and 2 times (depending on the viscosity of the fluid) when switching between a planar swimming induced by sinusoidal actuation and helical swimming induced by a conical magnetic field. This characteristic is also shown in \cite{chwang1971note} for self propelled swimmers. However, our work differs from these approaches as it does not rely on an a priori prescribed actuation pattern or shape deformation but optimizes the 3D driving magnetic field of the swimmer which allows the generation of swimmer-specific optimal actuation.
From figure \ref{Trajs_all},  we see that the dynamic model accurately predicts the horizontal displacements of the swimmer under both magnetic fields. It is less accurate in predicting the value of the amplitudes of the oscillations of the swimmer along the $y$ and $z$ axis as can be seen from comparing figures  \ref{fig_trajectory} and \ref{Displacement3D}, the shapes of the predicted and observed trajectories however match qualitatively.
The observed deformation patterns of the tail of the swimmer under a period of the sinusoidal and the optimal field are shown in Figure \ref{fig:shape_exp}. The optimal field induces a movement of the tail of the swimmer in both planes, in contrast with the sinusoidal field, which, expectedly, only deforms the tail of the swimmer in one plane. In the case of the optimal actuation, we see that the deformation of the tail of the swimmer is more significant in the top plane (see Figure \ref{fig:shape_exp} $\boldsymbol{(a)}$ and $\boldsymbol{(b)}$) compared to the side plane where we observe mainly a beating behavior, since the swimmer stays mostly straight in this plane (see Figure \ref{fig:shape_exp} $\boldsymbol{(c)}$ and $\boldsymbol{(d)}$).  
Although the main focus of this work was to optimize the speed of the flexible robot for swimming along a straight line, this swimming strategy is easily implementable for open loop or closed loop path following (see \cite{oulmas20173d} for an example) by applying the static component of the magnetic field in the direction tangent to the curve, and the time-varying components in the normal and binormal directions.

~\\
\section{Conclusion}
~\\
We have investigated the design of optimal actuation patterns for a flexible low Reynold number swimmer actuated by  magnetic fields using  a simple computationally inexpensive model that predicts the horizontal displacement of the swimmer. This provides an automated procedure for the optimal control design of  flexible magnetic low-Reynolds swimmers. Using this approach, we simulate magnetic fields that maximize the horizontal propulsion speed of the swimmer. From this, we are able to propose a novel magnetic actuation pattern that allows the swimmer to swim significantly faster compared to the usual sinusoidal actuation. This actuation pattern is experimentally validated on a flexible low Reynolds swimmer. The dynamic model used accurately predicts the horizontal displacement of the experimental swimmer under both the optimal and sinusoidal actuation. The simulation and experimental results show that it is necessary to go out-of plane in order to maximize the propulsion speed of flexible magnetic low-Reynolds swimmers. 
Although the dynamic model used is limited in its accuracy, since it is based on an approximation of the fluid-structure interaction of the swimmer, this study showcases the usefulness of such a simplified model for computationally-inexpensive control design for the actuation of flagellar magnetic swimmers. 
Current research is focused on finding a solution to allow propulsion of flagellated swimmers in confined environments (\cite{ISHIMOTO2014187,ShumGaffney15}) or in the presence of obstacles, such as the human vasculature (\cite{JangAho18}). For this purpose, the technique presented in this paper could be generalized to find  magnetic fields that allows the propulsion of the robot in a confined environnement by adapting the model.

\bibliographystyle{unsrt}

\end{document}